\documentclass{PoS}
\usepackage{amssymb}
\PoS{PoS(LAT2005)260}

\title{Unitary Evolution on a Discrete Phase Space }

\ShortTitle{Unitary Evolution on a Discrete Phase Space }

\author{Emmanuel G. Floratos\\
        Physics Department, University of Athens, Athens, Greece\\
        E-mail: \email{mflorato@phys.uoa.gr}}

\author{\speaker{Stam Nicolis}\\
        CNRS-LMPT Tours, Parc Grandmont, 37200 Tours, France\\
        E-mail: \email{stam.nicolis@lmpt.univ-tours.fr}}

\abstract{We construct unitary evolution operators on a phase space with 
power of two discretization. These operators realize the metaplectic
representation of the modular group $SL(2,{\mathbb{Z}}_{2^n})$. It acts 
in a natural way on the coordinates of the non-commutative 2-torus,
${\mathbb{T}}_{2^n}^2$  and thus is relevant for non-commutative field
theories as well as theories of quantum space-time. The class of operators may
also be useful for the efficient realization of new quantum algorithms.}

\FullConference{XXIIIrd International Symposium on Lattice Field Theory\\
		 25-30 July 2005\\
		 Trinity College, Dublin, Ireland}

\begin{document}

\section{Introduction}
Recent progress in M-theory indicates that spacetime itself becomes 
noncommutative at scales where D-branes play an important
role~\cite{matrixmodel,schwarz}.
This noncommutativity comes about in a rather natural way because
D-branes are charged, gravitational solitons, 
moving in backgrounds
with magnetic flux and their worldvolume acquires non-commutative geometry. 
What happens is analogous to the  Landau problem, where 
the noncommutativity of the two, real, space coordinates is brought
about by the magnetic flux\cite{connes}. The strength of the flux 
provides a measure of non-commutativity. This is the first time where explicit 
dynamics on non-commutative spacetime has been throughly studied and it has
led to a new understanding of  Yang--Mills theories (with fluxes) in {\em
  commutative} spacetime as $U(1)$ gauge theories in non-commutative
spacetime. 

The new insight is the trading of spacetime non-commutativity with the
non-commutativity of the gauge group. In general it has been established that
Yang--Mills  gauge theories (in commutative spacetime) have similar
short-distance behavior but different {\em long--distance} behavior from 
{\em non-commutative} Yang--Mills theories. 

The above facts led people to think again the old idea of discretization of
spacetime with cellular structure, which could be described by
non-commmutative coordinates. For Minkowski spacetimes care must be taken to
use non-commutativity only in the {\em space} part, otherwise unitarity in
perturbation theory is lost.   

More drastic ideas, not coming from D-brane studies, have been advanced by 't
Hooft\cite{hooft}
 in an effort to reconcile gravity with quantum mechanics. The proposal
is that, at scales where gravity becomes strong, it is no longer appropriate 
to work with continuous variables, but at the classical level, to use {\em
  discrete} labels for spacetime coordinates as well as the dynamical state of
the system. Quantum mechanics arises from the superoposition of these
classical dynamical states and the mapping of the configuration and momentum
space in wavefunctions on a finite--dimensional (and discrete) Hilbert
space. Locality in space is lost, since there is a minimal lengthscale and,
because of causality, locality is lost in time as well. Hamiltonians are thus
non-local and only unitary, one-time-step evolution operators have any
meaning. 

In this contribution we present the construction of unitary evolution
operators on a toroidal phase space with power of two discretiztion. This case 
could not be included in the previous cases of prime or odd discretization\cite{fqmdikamas}
due to the impossibility of defining 1/2 mod $2^n$. The solution we present
here relies in absorbing these factors and considering twice as many points
per phase space direction; We then show that the restriction to half the
points is consistent and present the transformation that groups them together
in  order to ensure a unitary evolution.

We study a  model for a discrete and periodic
one--dimensional space and, at the same time, discrete and periodic momentum,
i.e. a discretized, toroidal, phase space, ${\mathbb Z}_N\times {\mathbb Z}_N$, on
which classical, linear, maps, elements of $SL(2,{\mathbb Z}_N)$, which
discretize continuous, $SL(2,{\mathbb R})$ maps on the 2-torus, of unit length, 
${\mathbb T}^2$, for motions on the rational points, with denominator $N$. Taking
$N\to\infty$ a subset of possible trajectories has a smooth limit and we
recover $SL(2,{\mathbb R})$. This discretization has the nice property,
 that can be transferred to the quantum--mechanical level, 
of assigning to each 
${\sf A}\in SL(2,{\mathbb Z}_N)$, a unique, $N\times N$, unitary map, 
$U({\sf  A})$, which represents faithfully sequences of classical maps that
are irreducible representations of $SL(2,{\mathbb Z}_N)$. This can be done for
any $N$ odd using as elementary blocks the case $N=p^n$, $p$ prime and, for 
the general case, $N=p_1^{n_1}\times \cdots p_k^{n_k}\cdots $, we use the 
basic property of $SL(2,{\mathbb Z}_N)$,
$$
SL(2,{\mathbb Z}_N)=\otimes SL(2,{\mathbb Z}_{p_k^{n_k}})
$$
In the case $N=2^n$, or, more generally, $N$ even, there are constraints
imposing periodicity on the unitary matrices $U({\sf A})$, which restrict 
the    group $SL(2,{\mathbb Z}_N)$ to its normal subgroup 
$$
NSL(2,{\mathbb Z}_N)=
\left(\begin{array}{cc}
\mathrm{odd} & \mathrm{even}\\
\mathrm{even} & \mathrm{odd}
\end{array}
\right)
$$
There is an additional subset of $SL(2,{\mathbb Z}_N)$, consisting of elements
$$
{\sf A}=
\left(\begin{array}{cc}
\mathrm{even} & \mathrm{odd}\\
\mathrm{odd} & \mathrm{even}
\end{array}
\right)
$$
which, together with the elements of $NSL(2,{\mathbb Z}_N)$, form a bigger
subgroup, which satisfy the periodicity constraints. 

In the next section  we give the basic definitions for the metaplectic
representation and we quantize translations and dilatations. Using these 
as building blocks, it is thus possible to quantize any classical action. 
We close with a brief discussion of direction of future inquiry.

\section{Unitary Evolution for Translations and Dilatations}\label{TransDil}
We start with the observation that any matrix ${\sf A}$, element of 
$SL(2,{\mathbb{Z}}_{N})$, may be written as
\begin{equation}
\label{schurdecomp}
\underbrace{\left(\begin{array}{cc} a & b \\ c & d\end{array}\right)}_{ {\sf
    A} }=
\underbrace{\left(\begin{array}{cc} 1 & bd^{-1} \\ 0 & 1\end{array}\right)}_{
  {\sf L}(bd^{-1}) }
\underbrace{\left(\begin{array}{cc} d^{-1} & 0 \\ 0 & d\end{array}\right)}_{
  {\sf D}(d) }
\underbrace{\left(\begin{array}{cc} 1 & 0 \\ cd^{-1} & 1\end{array}\right)}_{
  {\sf R}(cd^{-1}) }
\end{equation}
i.e. in terms of translations and dilatations in phase space. We note that
obstructions to this construction may appear, if $d$ is not invertible mod
$N$. For $N=2^n$ this means, in particular, that $d$ should be odd. 
We shall try to construct the unitary operator, $U({\sf A})$ from the
corresponding unitary operators
$$
\begin{array}{ccc}
U^{\mathrm{L}}(x)\equiv U({\sf L}(x)) & U^{\mathrm{D}}(d)\equiv U({\sf D}) & 
U^{\mathrm{R}}(y)\equiv U({\sf R}(y))
\end{array}
$$
For any discretization $N$ it is possible to find operators $J_{r,s}$ that 
generate the Heisenberg--Weyl group. They are given in terms of the clock and
shift operators $Q$ and $P$ by 
\begin{equation}
\label{Jrs}
J_{r,s}=\omega_N^{rs/2}P^rQ^s
\end{equation}
where $\omega_N\equiv\exp(2\pi\mathrm{i}/N)$. 
In the basis where the clock operator $Q$ is diagonal the matrix elements are
given by 
\begin{equation}
\label{matrixelementsJPQ}
\begin{array}{ccc}
P_{k,l}=\delta_{k-1,l} & Q_{k,l}=\omega_N^k\delta_{k,l} &
\left[J_{r,s}\right]_{k,l}=\delta_{k-r,l}\omega_N^{s(k+l)/2}
\end{array}
\end{equation}
where $k,l=0,1,\ldots,2^n-1$. 
We already notice the $1/2$ factors that need to be defined seperately, when 
$N=2^n$. The unitary operators $U({\sf A})$ need to satisfy two requirements:
\begin{itemize}
\item They realize a group representation: for any two operators, 
${\sf A}, {\sf B}\in SL(2,{\mathbb{Z}}_{2^n})$ we have
\begin{equation}
\label{group}
U({\sf A}\cdot {\sf B})=U({\sf A})\cdot U({\sf B})
\end{equation}
\item They realize the {\em metaplectic} representation: for any point, 
$(r,s)$,  of the classical phase space
\begin{equation}
\label{metaplectic}
U({\sf A})J_{r,s}U({\sf A})^{-1}=J_{(r,s){\sf A}}
\end{equation}
\end{itemize}
In this section we shall sketch the construction for $U^{\mathrm{L}}(x)$, 
$U^{\mathrm{R}}(y)$ and $U^{\mathrm{D}}(d)$. Details may be found in the
paper\cite{FloratosNicolis22theN}. 

We first consider a larger space, of $2^{n+1}-1$ points,
i.e. $SL(2,{\mathbb{Z}}_{2^{n+1}})$. In this space we define the matrix
elements of $J_{r,s}$ as 
\begin{equation}
\label{Jrstwice}
\left[J_{r,s}\right]_{k,l}=\delta_{k-r,l}^{(n+1)}\omega_{n+2}^{s(k+l)}
\end{equation}
where $\omega_n\equiv\exp(2\pi\mathrm{i}/2^n)$ and the superscript on the
Kronecker delta indicates that the operation is performed mod $2^{n+1}$. Note
that we have absorbed the $1/2$ factor in the order of the root of unity. 

However the ``physical'' points correspond to the {\em even} values of the 
indices $r,s,k,l$ and, thus the ``physical'' sub-space is, indeed,
$2^n$--dimensional.  

It is possible to show that the following operators realize a group
representation and the metaplectic reprsentation:
\begin{itemize}
\item
  $$
\left[U^{\mathrm{L}}(x)\right]_{k,l}=\frac{1+(-)^k}{2}\omega_{n+2}^{xk^2}\delta_{k,l}^{(n+1)}$$
\item 
$$
\left[U^{\mathrm{D}}(d)\right]_{k,l}=\delta_{k,dl}^{(n+1)}
$$
\item 
$$
\left[U^{\mathrm{R}}(y)\right]_{k,l}=\left[F^{-1}U^{\mathrm{L}}(-y)F\right]_{k,l}=\frac{1}{2^{n+1}}\sum_{m=0}^{2^{n}-1}\frac{1+(-)^{l+k}}{2}
\omega_{n}^{-m^2x+m(l-k)}
$$
\end{itemize}
where $F$ is the Fourier Transform operator, 
$$
F_{k,l}=\frac{\omega_{n+1}^{kl}}{\sqrt{2^{n+1}}}
$$
Here the indices $k,l=0,1,\ldots,2^{n+1}-1$-however the projector 
$$
\frac{1+(-)^k}{2}
$$
projects on the even values, $k,l=0,2,4,\ldots,2^{n+1}-2$. Similarly, when
checking the metaplectic representation, eq.~(\ref{metaplectic}), the indices
$r,s=0,2,4, \ldots,2^{n+1}-2$. That this is consistent may be deduced from the 
fact that the points of the classical phase space
$(r=\mathrm{even},s=\mathrm{even})$ are transformed among themselves by a
matrix ${\sf A}\in SL(2,\mathbb{Z}_{2^n})$. We thus may check that the evolution
is unitary on this  sub-lattice. However, since we have introduced a projector, 
unitarity is spoiled on the original lattice. Remains then to construct an
operator that rearranges the sites in such a way as to render unitarity
manifest on the even sub-lattice. Such an operator is the {\em bit reversal}
operator used in the Fast Fourier Transform\cite{nr,FloratosNicolis22theN}! 

We thus may conclude that the general classical action ${\sf A}\in
SL(2,{\mathbb{Z}}_{2^n})$, with the element $d$ odd, may be consistently
quantized through an embedding in a space of twice as many points. Details and 
proofs may be found in \cite{FloratosNicolis22theN}. 
\section{Conclusions and Outlook}\label{Outlook}
We have presented the construction of unitary evolution operators that
consistently quantize the classical action on phase spaces of power of two
discretization. This completes the program of papers\cite{fqmdikamas} and
may open the way for new quantum algorithms based on operators that are more
general than the Fourier Transform. Furthermore the dimensionality considered
here may be useful in the description of systems with fermionic degrees of
freedom.

\end{document}